\documentclass[prb,superscriptaddress,twocolumn]{revtex4}
\usepackage{epsfig}
\begin{document}

\title{Accuracy of the Hartree-Fock method for
Wigner molecules at high magnetic fields}
\author{B. Szafran}
\affiliation{Departement Natuurkunde, Universiteit Antwerpen
(UIA), B-2610 Antwerpen, Belgium} \affiliation{Faculty of Physics
and Nuclear Techniques, AGH University of Science and Technology,
Krak\'ow, Poland}
 \author{S. Bednarek}
 \affiliation{Faculty of Physics and Nuclear Techniques, AGH
University of Science and Technology, Krak\'ow, Poland}
 \author{J. Adamowski}
\affiliation{Faculty of Physics and Nuclear Techniques, AGH
University of Science and Technology, Krak\'ow, Poland}
\author{M.B. Tavernier}
\affiliation{Departement Natuurkunde, Universiteit Antwerpen
(UIA), B-2610 Antwerpen, Belgium}
\author{Egidijus Anisimovas}
\affiliation{Departement Natuurkunde, Universiteit Antwerpen
(UIA), B-2610 Antwerpen, Belgium}
\author{F.M. Peeters}
\affiliation{Departement Natuurkunde, Universiteit Antwerpen
(UIA), B-2610 Antwerpen, Belgium}

 \begin{abstract} Few-electron systems confined in two-dimensional
parabolic quantum dots at high magnetic fields are studied by the
Hartree-Fock (HF) and exact diagonalization methods. A generalized
multicenter Gaussian basis is proposed in the HF method. A
comparison of the HF and exact results allows us to discuss the
relevance of the symmetry of the charge density distribution for
the accuracy of the HF method. It is shown that the energy
estimates obtained with the broken-symmetry HF wave functions
become exact in the infinite magnetic-field limit.
 In this limit
the charge density of the broken-symmetry solution can be
identified with the classical charge distribution.

\end{abstract}
\maketitle

\section{Introduction}

Properties of electron systems confined in quantum dots at high
magnetic fields have recently become a subject of intensive
theoretical studies
\cite{McDo1,Koonin,Yan1,Yan2,Sza03,Sza032,Kainz,Reusch,Rei1,Reir,makchak,Eto,Imma,Mani,Matulis,Mikha,McDo2,Wojs,Dine,Sza033,Rontani,Peet}.
These studies were inspired on the one hand by the experimental
investigation of the addition spectra of vertical quantum dots
\cite{Oster} at high magnetic fields, which revealed a rich
structure of magnetic-field induced ground-state transformations
in the confined electron system
\cite{McDo1,Koonin,Yan1,Yan2,Sza03,Sza032,Kainz,Reusch,Rei1,Reir,makchak,Eto,Imma,Mani,Matulis,Mikha,McDo2,Wojs,Dine,Sza033,Rontani,Peet,Oster,SBA02,Maksym,bp2,bp3},
and on the other hand by the search for a new symmetry in
few-electron systems. One of the most interesting problems in this
research is the possibility of the formation of Wigner molecules
\cite{Koonin,Yan1,Yan2,Sza03,Sza032,Reusch,Kainz,Rei1,Reir,Maksym,bp2,bp3}
in which the confined electrons are distinctly spatially
separated. Previous theoretical studies are based on the
Hartree-Fock (HF) method
\cite{Koonin,Yan1,Yan2,Sza03,Sza032,Reusch}, density functional
theory (DFT) \cite{Rei1}, and the exact diagonalization (ED)
scheme
\cite{makchak,Eto,Imma,Mani,Matulis,Mikha,McDo2,Wojs,Dine,Sza033,Rontani}.
The model confinement potential used most commonly
\cite{McDo1,Koonin,Yan1,Yan2,Sza03,Sza032,Reusch,Kainz,Rei1,Reir,makchak,Eto,Imma,Mani,Matulis,Mikha,McDo2,Wojs,Dine}
is the two-dimensional (2D) cylindrically symmetric harmonic
oscillator potential, which is a reasonable approximation of the
confinement potential \cite{BSA} in vertical quantum dots
\cite{Oster}.

The external magnetic field induces ground-state transformations
in the quantum-dot confined $N$-electron system, which are
associated with changes of the total angular momentum and the
total spin. At a certain magnetic field the electrons become spin
polarized and occupy the lowest-energy Fock-Darwin states with the
$z$ component of the angular momentum changing from 0 to $(1-N)
\hbar$ \cite{Reir}. This state is called a maximum density droplet
(MDD) \cite{McDo1}. In the MDD the $z$ components of the total
spin and total angular momentum take on the absolute values
$N\hbar/2$ and $N(N-1)\hbar/2$, respectively. The MDD ground state
is predicted by the ED method \cite{Mikha,McDo2,Wojs,Dine,Sza033}
as well as by the HF \cite{Koonin} and DFT \cite{Rei1} methods.
Higher magnetic fields lead to a decay of the MDD, which has been
observed experimentally \cite{Oster}. This decay, considered in
the framework of the ED method \cite{Mikha,McDo2}, is related to
the increase of the absolute value of the total angular momentum
above the value corresponding to the MDD phase. The unrestricted
HF \cite{Koonin,Yan1,Yan2,Sza03} and DFT \cite{Rei1} methods
predict that the ground state of the electron phase created after
the decay of the MDD possesses a charge density which does not
reproduce the symmetry of the external confinement potential.
These states will be called ''broken-symmetry states'' throughout
the present paper. In the broken-symmetry solutions obtained by
the HF and DFT methods, the electrons become localized at separate
space sites forming a Wigner molecule
\cite{Koonin,Sza03,Sza032,Kainz,Rei1,Reir,Maksym,bp2,bp3}. In the
ED method the separation of the electrons, i.e., the formation of
the Wigner molecules, appears in the relative coordinates of the
electron system and is not necessarily related to the broken
symmetry of the charge density \cite{Reir}. The HF broken-symmetry
solutions are not eigenfunctions of the angular momentum operator,
but are degenerate with respect to rotations, i.e., can be
oriented at an arbitrary angle. The rotational symmetry of the HF
broken-symmetry solutions for the few-electron system can be
restored with a post-HF treatment \cite{Yan2}.

A multicenter basis with the one-electron wave functions was used
in the theoretical studies of the 2D Wigner crystals
\cite{WC1,WC2}. Recently, a similar multicenter basis was used
\cite{Yan2,Sza03,Sza032,Kainz} to study the quantum-dot confined
electron systems. The papers \cite{Yan2,Sza03,Sza032,Kainz} were
based on the unrestricted HF method \cite{Yan2,Sza03,Sza032} and
the variational method \cite{Kainz} with the trial wave function
in the form of a single Slater determinant with non-orthogonal
one-electron wave functions. These one-electron wave functions
\cite{Yan2,Sza03,Sza032,Kainz} at high magnetic field yield
point-like charge density distributions. Therefore, in the limit
of infinite magnetic field, the ground-state charge density,
obtained with the multicenter basis, is identical with the
lowest-energy classical configuration of the point charges
\cite{C1,Pe1,Pe2}. The multicenter basis is a very efficient tool
for the investigation of the Wigner molecules, since it requires
only a single basis function per electron, while the convergence
of the HF energy estimates in the one-center Fock-Darwin basis
\cite{Koonin} is very slow in the regime of the island-like Wigner
localization. Using the unrestricted HF method with the
multicenter basis (MCHF),  Szafran \emph{et al.} \cite{Sza03} have
shown that the external magnetic field leads to transformations of
the ground-state symmetry of the Wigner molecules. Only in the
high magnetic field limit the ground-state phase (isomer of the
Wigner molecule) \cite{Sza03} corresponds to the configuration of
electrons, which is identical with that of a classical system of
point charges \cite{C1,Pe1,Pe2}.

In the present paper, we perform a detailed study of the physics
behind the formation of the Wigner molecules. In particular, we
discuss the accuracy of the broken symmetry solutions obtained
with the multicenter basis in comparison with the ED results for
two, three, and four electrons. The physical interest of this
study relies on the investigation of the quantum systems in a
classical localization limit.

The paper is organized as follows: in Section II we briefly
describe the theoretical methods, in Section III we present our
numerical results, in Section IV -- the discussion, and in Section
V -- the conclusions and the summary. In Appendix, we derive the
wave function used in the present calculations.

\section{Theory}

We consider the $N$-electron system confined in the 2D harmonic
oscillator potential with frequency $\omega_0$, subject to the
external magnetic field $B$ oriented perpendicularly to the
quantum dot plane \cite{Sza03,Sza032}. We apply the MCHF and ED
methods. In the MCHF method we assume that all the electrons are
spin polarized by the magnetic field and apply the Landau gauge,
i.e., $\mathbf{A}=(-By,0,0)$. We expand one-electron wave function
$\Psi_\mu(\mathbf{r})$ of the $\mu$-th occupied state
($\mu=1,\ldots,N$)
\begin{equation}
\Psi_\mu(\mathbf{r}) = \sum_{i=1}^{N} c_i^\mu \psi_{\mathbf{R}_i}
(\mathbf{r}) \;, \label{expansion}
\end{equation}
in the basis
\begin{eqnarray}
\psi_{\mathbf R}({\mathbf r})= &(\alpha/2\pi)^{1/2} \exp\{-(\alpha
/4) \left({\mathbf r} -{\mathbf R}\right)^2 + \nonumber \\ &
(i\beta/2)(x-X)(y+Y)\} \;, \label{wv}
\end{eqnarray}
where ${\mathbf R}=(X,Y)$, ${\mathbf r}=(x,y)$, and $\alpha$ and
$\beta$ are treated as nonlinear variational parameters. Function
(\ref{wv}) with $\alpha=\beta=m^*\omega_c/\hbar$ is the wave
function of the lowest Landau level ($\omega_c=eB/m^*$ is the
cyclotron frequency and $m^*$ is the electron effective mass).
Moreover, function (\ref{wv}) with
$\alpha=(2m^*/\hbar)\sqrt{\omega_0^2+\omega_c^2/4}$ and
$\beta=m^*\omega_c/\hbar=eB/\hbar$ is the eigenfunction of the
Fock-Darwin ground state for the lateral parabolic confinement
potential centered at point ${\mathbf R}$ with the energy equal to
${\hbar}\sqrt{\omega_0^2+\omega_c^2/4}$ (see Appendix).  The
probability density associated with wave function (\ref{wv}) is
the Gaussian centered around point ${\mathbf R}$. The centers of
basis functions (\ref{wv}) are taken from scaled configurations
$\{\mathbf{R}^{class}\}$ of classical Wigner molecules, i.e.,
$\{\mathbf{R}\} \equiv \mathbf{R}_i=\sigma\mathbf{R}^{class}_i$,
where $i=1, \ldots, N$ \cite{Sza03,Sza032}. The scaling parameter
$\sigma$ is the third nonlinear variational parameter used in the
present approach. In Refs. [5,6] only two nonlinear variational
parameters were used, namely, the scaling parameter $\sigma$ and a
single variational parameter $\alpha=\beta$. In the following, we
will show that the introduction of the two independent variational
parameters in the real and imaginary part of the exponent in
Eq.(\ref{wv}) leads to a significant improvement of the
variational energy estimates at finite magnetic field. Throughout
the present paper, the basis with the restriction $\alpha=\beta$
will be referred to as ''restricted'' and without it as
''generalized''.

In the present paper we also apply the ED method. The ED results,
that in principle are the exact solutions of the few-electron
Schr\"odinger equation, are used as reference data for the
estimation of the accuracy of the MCHF results. In the ED
calculations we use the symmetric gauge [${\bf A}
=(-By/2,Bx/2,0)$], that allows us to exploit the angular symmetry
of the one-electron wave functions. The Schr\"odinger equation for
the $N$-electron system can be separated into the center-of-mass
and relative-coordinate equations \cite{Matulis,Dine,Peet}. The
center-of-mass eigenproblem possesses an analytical solution. For
$N=2$ the relative-motion eigenproblem can be easily solved
numerically with an arbitrary precision using a one-dimensional
finite difference method.  This approach \cite{Matulis,Dine}, used
in the present paper for the two-electron system, is not
applicable to the systems with a larger number of electrons. Thus,
for $N=3$ and $N=4$ the ED procedure is constructed according to
the configuration interaction method. First, we solve the
Schr\"odinger equation for a single noninteracting electron with a
definite angular momentum using a finite-difference approach on
the one-dimensional mesh with 200 points. Next, we use the
single-electron wave functions to construct Slater determinants
with the required total angular momentum and spin. The
$N$-electron Schr\"odinger equation is diagonalized in the
orthonormal basis of Slater determinants with proper spin-orbital
symmetry and the Coulomb matrix elements are integrated
numerically. The basis, i.e., the choice of the single-electron
wave functions forming the Slater determinants, is optimized
separately for each state. The calculations have been performed
with a precision better than 0.01 meV. For $N=4$ this precision
requires the application of the basis containing up to 2000 Slater
determinants. The ED calculations of the ground state for $N=3$
and 4 are carried out up to 20 T with the maximal absolute values
of the angular momentum equal to 18 and 34 $\hbar$, respectively.

\begin{figure}[htbp]{\epsfxsize=75mm
                \epsfbox[40 215 535 747]{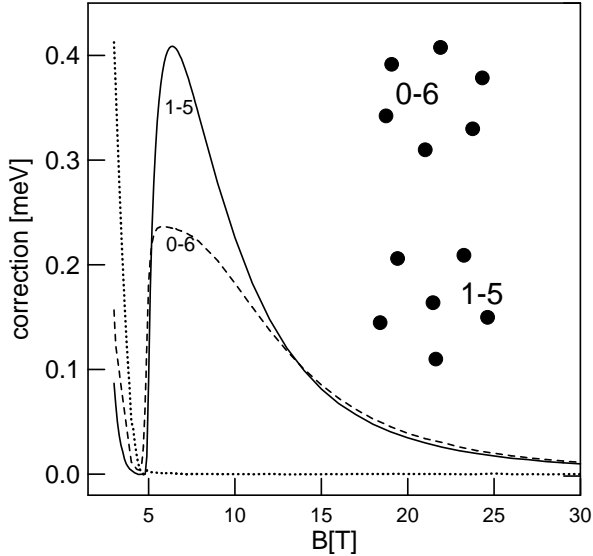}}\newline
\caption{Correction to the energy estimates obtained with the
generalized MCHF method and calculated with respect to the results
of the restricted HF method ($\alpha=\beta$) for the six-electron
system with the configuration 1-5 (solid line) and 0-6 (dashed
line) as a function of magnetic field $B$. The dotted line shows
the overestimation of the energy obtained for 1-5 configuration
with the value of $\beta$ fixed at $eB/\hbar$. Inset: the
classical 0-6 and 1-5 Wigner molecules. \label{pop}}
\end{figure}

\section{Results}

Let us first discuss the corrections to the MCHF energy estimates
obtained with the generalized wave function (\ref{wv}). We use as
an example the system of six confined electrons. Figure  \ref{pop}
shows the difference between the energy estimates obtained with
and without the restriction $\alpha=\beta$ in wave function
(\ref{wv}). In the calculations we use the material parameters of
GaAs, i.e. $m^*=0.067$ $m_e$, dielectric constant
$\varepsilon=12.9$, the effective Land\`e factor $g^*=-0.44$, and
assume the confinement energy $\hbar\omega_0=3$ meV
\cite{Koonin,Sza03}. The centers of basis (\ref{expansion}) are
taken from scaled classical configurations 1-5 and 0-6 [the phases
(isomers) of the Wigner molecule are labelled by the numbers of
electrons localized in the subsequent rings starting from the
innermost one]. Phase 0-6 is the one, in which the Wigner molecule
is created \cite{Sza03,Reir,Mani} after the MDD breakdown, and
possesses an intermediate character.
\begin{figure}[htbp]{\epsfxsize=77mm
                \epsfbox[20 30 592 573]{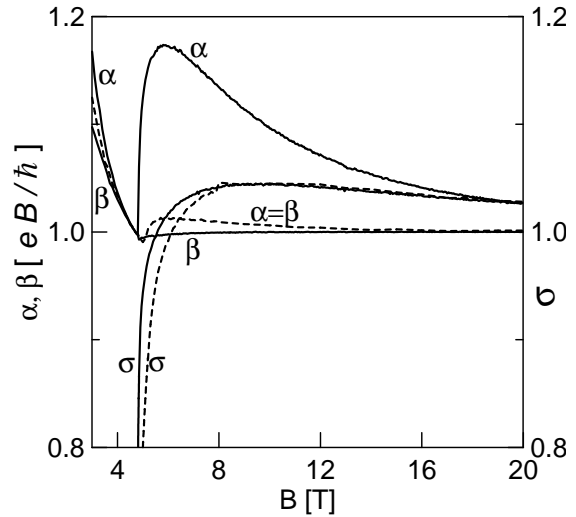}}\newline
\caption{Optimal values of the variational parameters for the 1-5
configuration obtained with the generalized ($\alpha\neq\beta$,
solid lines) and restricted  ($\alpha=\beta$, dashed lines) basis.
Parameters $\alpha$ and $\beta$ are expressed in units $eB/\hbar$,
parameter $\sigma$ is dimensionless. \label{parwar}}
\end{figure}
The 1-5 configuration is the lowest-energy configuration of the
classical Wigner molecule \cite{Pe1}. This is also the
ground-state configuration of the six-electron quantum system at
high magnetic field \cite{Sza03}. In the magnetic field below 4.9
T the mutlicenter bases with both the 0-6 and 1-5 configurations
mimic the cylindrically symmetric MDD charge distribution
\cite{Sza03}. The correction obtained with the generalized basis
possesses a minimum in the magnetic field inducing the MDD
breakdown, for which, the application of the generalized basis
does not improve the results. However, the generalized basis leads
to considerable corrections to the energy both in the MDD
stability regime and after the MDD breakdown. This correction
falls down to zero in the high-magnetic field, in which both the
generalized and restricted bases work with nearly the same
precision.

The dotted line in Fig. \ref{pop} shows the overestimation of the
energy obtained for 1-5 configuration with fixed value of
parameter $\beta=eB/\hbar$. The value of $\beta$ has an influence
on the MCHF charge density due to the interference of the
single-electron wave functions [Eqs.~(\ref{expansion},\ref{wv})]
centered at different sites. Fig. \ref{pop} shows that the
variation of $\beta$ has a large influence on the estimates of the
energy in the MDD phase. On the other hand, in the Wigner molecule
phase, the value of $\beta$ can be safely put equal to $eB/\hbar$.

Figure \ref{parwar} shows the magnetic-field dependence of the
optimal variational parameters obtained for the 1-5 configuration
in expansion (\ref{expansion}). The dependence of the scaling
parameter $\sigma$ is qualitatively the same for both wave
functions. This parameter grows rapidly when the MDD breaks down
into the molecular phase. Generalized wave function (\ref{wv})
with $\alpha\neq\beta$ leads to the MDD decay at lower values of
the magnetic field, which is visible in the dependence of $\sigma$
on the magnetic field. Parameter $\sigma$ tends to 1 at high
magnetic fields for which the quantum Wigner molecule takes the
shape and size of its classical analog, i.e., $\{\mathbf{R}\}
\longrightarrow \{\mathbf{R}^{class}\}$. Parameters $\alpha$ and
$\beta$ decrease with the increasing magnetic field in the MDD
regime (cf. solid lines for $B<\sim 5$ T in Figure \ref{parwar}).
Just before the MDD breakdown $\alpha$ and $\beta$ take on very
close values, which leads to the minimal overestimation of the
energy obtained with restriction $\alpha=\beta$ at the MDD
breakdown (cf. Fig. \ref{pop}). We have found that the increase of
$\beta$ above $eB/\hbar$ in the MDD regime makes the local maxima
of the charge density 'sink' in the global flat maximum
characteristic \cite{Sza032} for the MDD phase. After the MDD
breakdown parameter $\beta$ quickly reaches $eB/\hbar$, i.e. takes
on the value which corresponds to both the lowest Landau and
Fock-Darwin levels [cf. the discussion of wave function (\ref{wv})
in Sec. II]. At high magnetic fields the wave functions (\ref{wv})
centered around different sites stop to overlap; in consequence,
$\beta$ stops to influence the MCHF charge density and takes the
value $eB/\hbar$, which ensures the equivalence of the centers of
Landau orbitals. The present finding that the parameter $\beta$
becomes equal to $eB/\hbar$ just after the MDD breakdown, in spite
of the non-vanishing overlap is not evident a priori. We can give
the following physical interpretation to this finding: the
electrons in the Wigner molecule behave as if they occupied the
independent one-particle Fock-Darwin ground-state orbitals, each
of them centered around its own local minimum of the potential
energy. The optimal value of $\alpha$ rapidly grows after the MDD
breakdown (cf. Fig. 2), which leads to the lowering of the energy
obtained in the Wigner molecule regime ($B>5$ T). The increase of
$\alpha$ above $eB/\hbar$ enhances the electron localization and
lowers the electron-electron interaction energy.

\begin{figure}[htbp]{\epsfxsize=75mm
                \epsfbox[4 154 594 672]{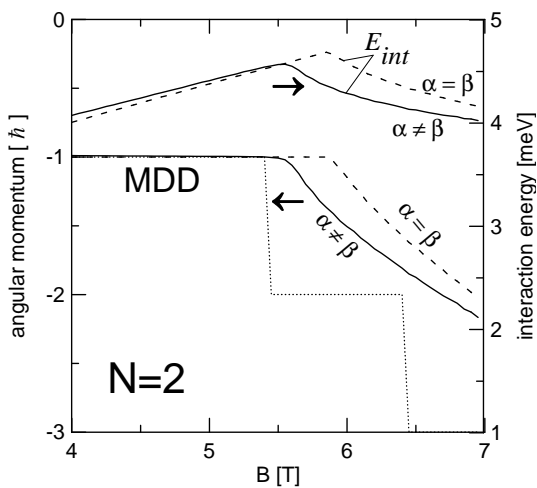}}\newline
\caption{Total angular momentum (left scale) of the exact ground
state of the two-electron system (dotted line) and MCHF
expectation values calculated with the generalized
($\alpha\neq\beta$, solid line) and the restricted
($\alpha=\beta$, dashed line) basis. The two curves marked by
$E_{int}$ show the expectation value of the electron-electron
interaction energy (right scale) calculated with the MCHF
methods.\label{2emdd}}
\end{figure}

\begin{figure}[htbp]{\epsfxsize=77mm
                \epsfbox[5 192 585 700]{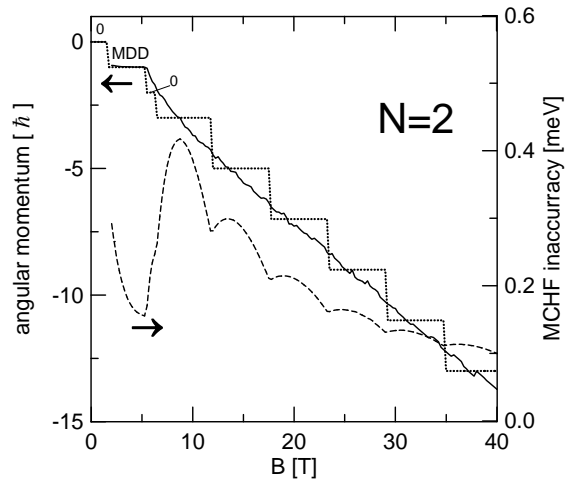}}\newline
\caption{Total angular momentum (left scale) of the exact ground
state of the two-electron system (dotted line) and the MCHF
expectation value calculated with the generalized basis (solid
line) as functions of magnetic field $B$ . The states with
unpolarized spins are marked by ''0''. The dashed line shows the
difference of the ground-state energy obtained with the MCHF and
the exact energy calculated with the ED scheme (right scale).
\label{m2}
 }
\end{figure}

Figure  \ref{2emdd} illustrates the MDD decay picture obtained
with the ED and MCHF methods for the two-electron system. The
dotted line shows the exact values of the $z$ component of the
total angular momentum as a function of the magnetic field. The
decay of the MDD is related with a stepwise decrease of the
angular momentum from $-\hbar$ to $-2\hbar$, which appears for
$B=5.45$ T. The solid (dashed) line shows the expectation value of
the total angular momentum obtained with the MCHF method using the
generalized (restricted) basis. Figure 3 shows that within the MDD
stability regime both the present MCHF methods reproduce the exact
value of the angular momentum. When the MDD breaks down, the MCHF
expectation value of the angular momentum decreases monotonically
with the increasing $B$, in contrast to the exact stepwise
behavior. The MDD breakdown obtained by the HF method is also
related to a cusp of the interaction energy \cite{Sza032} (cf. two
upper curves in Fig. \ref{2emdd}). In the MDD regime the charge
density distribution shrinks with increasing magnetic field, which
results in an increase of the interaction energy. The
transformation of the charge density from the droplet into the
molecular phase occurs when the interaction energy exceeds some
threshold value. The MCHF method with the generalized (restricted)
basis leads to a MDD breakdown for B=5.55 T (5.85 T).

Figure  \ref{m2} displays the exact ground-state angular momentum
for the two-electron system (dotted line) and the expectation
value obtained within the HF method with the generalized
multicenter basis (solid line) as well as the difference between
the MCHF  and the exact energy (dashed line). The $z$ component of
the total spin is equal to $\hbar$ with the exception of the
low-magnetic field ground state and the state which appears just
after the MDD breakdown. These two states possess zero spin and
are labelled by ''0'' in Figure \ref{m2}. Notice that the
expectation value of the total angular momentum follows quite well
the exact value. Moreover, the overestimation of the ground state
energy within the MCHF method decreases with increasing magnetic
field. This decrease is non-monotonous, and the MCHF inaccuracy
exhibit local minima for magnetic fields for which the exact
angular momentum changes.
\begin{figure}[htbp]{\epsfxsize=70mm
                \epsfbox[25 5 554 530]{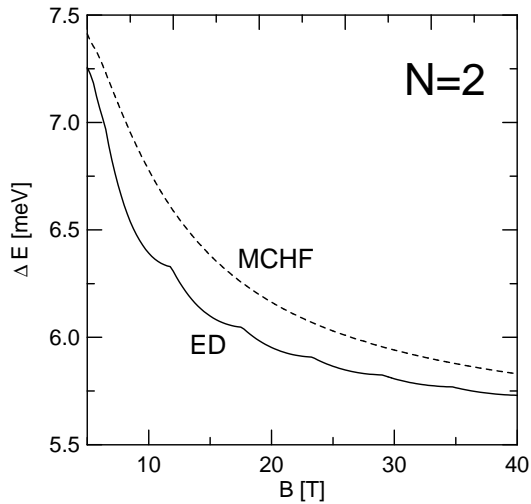}}\newline
\caption{Energy $\Delta E$ of the two-electron system calculated
with respect to the lowest Landau level as a function of magnetic
field $B$. Solid (dashed) curve show the exact (MCHF) results. }
\label{m2o}
\end{figure}

\begin{figure}[htbp]{\epsfxsize=78mm
                \epsfbox[23 200 591 679]{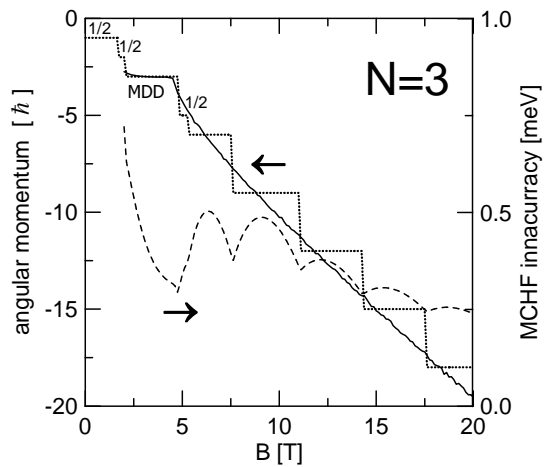}}\newline
\caption{Same as Figure \ 4 but now for the three-electron system.
The states with unpolarized spins are marked with ''1/2''.
 } \label{m3}
\end{figure}

The origin of this oscillatory behavior is explained in Fig.
\ref{m2o}, which shows the MCHF energy estimate and the exact
ground-state energy of the two-electron system calculated with
respect to the lowest Landau level, i.e., $\Delta E=
E-2\times(\hbar \omega_c/2+ \hbar g^*\mu_BB/2$). The MCHF estimate
is a smooth function of the magnetic field, while the exact energy
possesses cusps at the magnetic fields at which the ground-state
angular momentum changes abruptly. For these magnetic fields the
MCHF estimate is visibly closer to the exact energy value, which
explains the local minima in Figure 4.

A similar comparative study between the exact and the MCHF results
has been made for the $N=3$ and $N=4$ systems. For the system of
three electrons the results are shown in Fig. 6, in which the
ground states with the total spin equal to $\hbar/2$ are labelled
by ''1/2''. The other states are fully spin polarized. Similarly
as in the case of the two-electron system the ground state which
appears after the MDD breakdown is not spin polarized. The
overestimation of the total energy obtained with the MCHF method
exhibits a similar qualitative dependence on the magnetic field as
for two electrons. In contrast to the two-electron case, the MCHF
method with the generalized basis predicts a breakdown of the MDD
for a slightly smaller magnetic field value ($B=4.6$ T) than the
exact result ($B=4.8$ T). The MCHF with the restricted basis
yields the magnetic field inducing the MDD breakdown $B=5$ T.

\begin{figure}[htbp]{\epsfxsize=80mm
                \epsfbox[6 183 580 635]{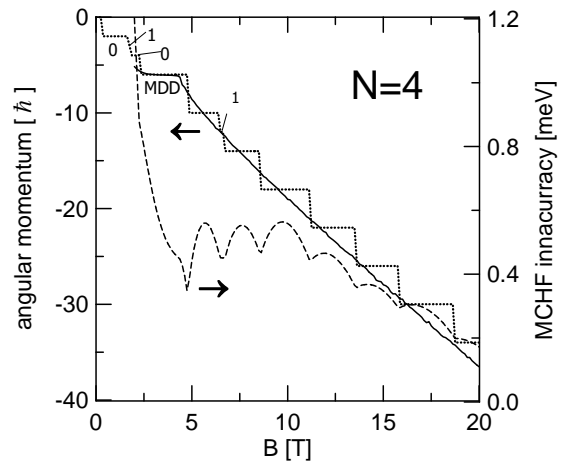}}\newline
\caption{Same as Figure \ 4 but now for the four-electron system.
The states with unpolarized spins are marked with the value of the
$z$-component of the total spin in $\hbar$ units.
 } \label{m4}
\end{figure}

\begin{figure}[htbp]{\epsfxsize=70mm
                \epsfbox[15 9 562 525]{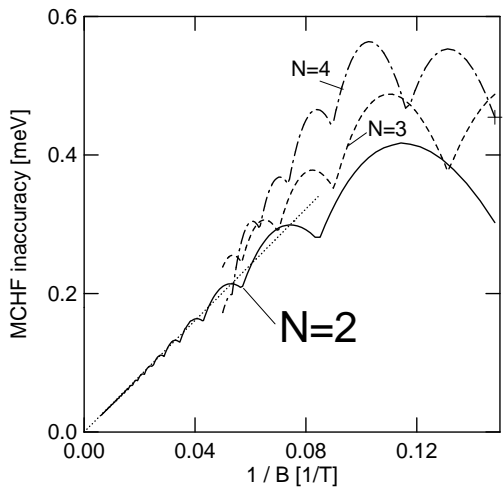}}\newline
\caption{The MCHF inaccuracy for the ground state for the
two-electron system as a function of $1/B$ for $N$=2 (solid line),
3 (dashed line) and 4 (dash-dotted line) The dotted line shows the
high-magnetic field asymptote for $N=2$, parametrized as $4.026/B$
[meV/T]. \label{assi}
 }
\end{figure}

A similar result for the four-electron quantum dot is shown in
Figure \ref{m4}. The non-fully-polarized ground states are marked
by the quantum numbers of the total-spin $z$ component ''0'' and
''1''. Contrary to the two and three electron systems, the
four-electron MDD decays into a spin-polarized state, but a
low-spin state still appears at the higher magnetic field.
The transitional appearance of low-spin ground states at magnetic
fields after the MDD breakdown have been reported first in Ref.
\cite{makchak} for $N=2$ and $N=4$.
The exact magnetic field inducing the MDD decay is equal to 4.75
T, while the MCHF with the generalized basis predicts a value of
4.38 T and the MCHF with the restricted basis gives 4.8 T. In this
case the MCHF with the restricted basis gives accidentally a
better estimate for the MDD breakdown field. For five electrons
the difference between both the MCHF estimates of the magnetic
field inducing the MDD decay is 0.4 T. For six and more electrons
the differences are not larger than 0.2 T.

Figures 4, 6, and 7 show that the MCHF inaccuracy decreases with
the increasing magnetic field. In order to find the high-field
asymptotic behavior of the MCHF energy estimate we have plotted in
Figure \ref{assi} the MCHF error as a function of $1/B$ for $N$ =
2, 3, and 4. The plot for two electrons covers the magnetic fields
up to 160 T, while the plots for three and four electrons are
drawn up to 20 T only. The results for $N=2$ show that at high
magnetic fields the MCHF inaccuracy is proportional to $1/B$. At
high magnetic fields the asymptotic behavior of the MCHF
inaccuracy for $N=2$ can be very well approximated by the function
$f(B)=4.026/B$ [meV/T], and consequently the MCHF approach becomes
exact for $B\rightarrow\infty$. The plots for $N=3$ and 4 in the
studied (narrower) range of the magnetic field exhibit a similar
tendency as that for $N=2$, however, they do not become linear
functions of $1/B$ for $B<20$ T. The comparison of the MCHF
inaccuracies for $N=2,3$, and 4 indicates that the overestimation
of the MCHF ground-state energy in the magnetic field range above
the MDD instability does not substantially increase with $N$. For
$B=20$ T the overestimation of the ground-state energy for
$N=2,3$, and 4 is equal to 0.21, 0.24, and 0.18 meV, respectively.

\section{Discussion}

Figs. \ref{m2}, \ref{m3} and \ref{m4} show that at high magnetic
field the exact ground-state angular momentum take the so-called
magic values \cite{Maksym,magic0,magic1,magic2} and change by
$N\hbar$. Only for these magic values of the angular momenta the
classical symmetry can be reproduced in the inner coordinates of
the quantum systems \cite{Maksym}. On the other hand the classical
symmetry is ensured in the MCHF by the present choice of centers
of orbitals (\ref{wv}) and the linear change of the expectation
value of the total angular momentum is related with the growing
localization of wave functions (\ref{wv}).

The results presented in Figs. \ref{m2}, \ref{m3},  \ref{m4}, and
\ref{assi} show that the broken-symmetry solutions obtained with
the MCHF method provide exact energy results in the high magnetic
field limit. This fact might be rather surprising, since one could
expect that the exact solutions of the few-electron Schr\"odinger
equation should also be the eigenfunctions of the angular momentum
operator. The exact eigenfunctions yield the charge density
distribution, which reproduces the symmetry of the confinement
potential. This apparent contradiction can be solved if we
consider the Schr\"odinger equation for the electron system
confined in the parabolic potential. This equation can be
separated in the center-of-mass and relative coordinates. In the
framework of the ED approach, the separation of the electrons,
i.e., the Wigner localization, appears in the relative (inner)
coordinates of the electron system, while the charge density in
the laboratory frame is affected by the center-of-mass motion. The
center-of-mass eigenproblem of the few-electron system has the
same form as the Fock-Darwin equation for the single electron. In
the high magnetic field limit, this equation possesses a
degenerate ground state (the lowest Landau state), for which the
angular momentum is arbitrary. In the case of this degeneracy a
superposition of ground states with different angular momenta is
still the ground state of the Fock-Darwin equation, even though it
leads to solutions with the broken symmetry of the charge density
distribution in the laboratory frame. In the HF method, the
formation of the Wigner molecule, i.e., the separation of the
electrons, is only possible in the broken-symmetry solutions. The
results of the present paper show that exact energy is obtained
for these broken-symmetry solutions in the infinite magnetic field
limit.

In the recent paper of Bednarek \emph{et al.} \cite{Chwiej} a
study of the accuracy of the HF method has been presented for the
quasi-one-dimensional (1D) structure. In the quasi-1D structures
the unrestricted HF method becomes exact \cite{Chwiej} in the
large quantum dots, in which the Wigner molecules are formed
\cite{Chwiej}. In the 1D structures, the HF method conserves the
two-particle parity symmetry of the exact solution; so, there is
no problem with the broken symmetry of the HF solutions like in
the 2D circularly symmetric quantum dots.

The present results show that the overestimation of the exact
energy obtained in the broken-symmetry MCHF solutions is
relatively small at the magnetic fields, for which the ground
state is degenerate. The ground state of the few-electron system
is twofold degenerate at these magnetic fields, which induce a
stepwise change of the ground-state angular momentum (cf. Figs.
\ref{m2}, \ref{m3}, and \ref{m4}). In this case, the exact ground
state can be a superposition of two states with different angular
momenta and therefore the charge density can have the symmetry
different from that of the external potential. For these fields
the broken symmetry of the HF solutions leads to the decrease of
the energy separation between the MCHF and ED results (cf. Fig.\
6),  which in turn causes the characteristic oscillations of the
MCHF error as a function of the magnetic field as shown in Figs.
\ref{m2}-\ref{m4}.

The application of the generalized variational basis results in a
modification of the phase diagram \cite{Sza03}, for the Wigner
molecules. This modification is due to the different precision of
the restricted wave function for different phases (cf. Fig. 1).
The critical magnetic fields for the MDD breakdown, obtained with
the generalized basis, are shifted toward lower values (cf. Fig.
3) and the range of the stability of different phases is modified.
The improved results conserve the characteristic features of the
original phase diagram \cite{Sza03} i.e., the intermediate phases
correspond to the configurations, for which a larger number of
electrons is gathered on the outer ring of the molecule in
comparison with the classical, high-magnetic-field, ground-state
configuration.

The present calculations performed for small number of electrons
indicate that in the high magnetic field the inner charge
distribution of electrons can be derived from classical
calculations for particles interacting via a Coulomb ($1/r$)
interaction. On the other hand in the Laughlin wave function
\cite{Laugh}, which seems to become an exact description of the
many-body state at high magnetic field, the distribution of
electrons in the inner coordinates corresponds to classical
configuration of particles interacting with a logarithmic
potential. It is therefore not excluded that for larger number of
electrons the localization may be different than for
electrostatically interacting classical particles.

\section{Conclusions and Summary}

We have investigated the quantum-dot confined $N$-electron system
at high magnetic fields using the HF method with the generalized
multicenter basis and the exact diagonalization method. We have
indicated that the magnetic-field dependence of the variational
parameters of the generalized basis can be used as one of the
signatures of the liquid-solid phase transition, i.e., the
breakdown of the MDD into the molecular phase. The occurrence of
the cusp of the interaction energy as a function of the magnetic
field and the decrease of the MCHF angular-momentum expectation
value below that corresponding to the MDD are the other signatures
of the MDD decay. We have discussed the accuracy of the energy
estimates obtained with the broken-symmetry HF solutions, which --
in contrast to the exact solutions of the Schr\"odigner equation
-- are not eigenfunctions of the total angular-momentum operator.
It turns out that the angular-momentum eigenvalues in the MDD
phase are reproduced with the high accuracy by the MCHF
expectation values, which at higher magnetic fields linearly
decrease with increasing $B$ in contrast to the exact stepwise
decrease. The results of the present paper show that the MCHF
inaccuracy decreases with the increasing magnetic field and that
the MCHF method basis yields the exact ground-state energy in the
infinite magnetic-field limit. We have found the characteristic
oscillations of the HF inaccuracy, which exhibits local minima at
those magnetic fields, for which the exact ground state is
degenerate with respect to the angular momentum. The relation of
these oscillations with the existence of the exact ground states
with the broken symmetry has been pointed out. The results of the
present paper show that the envelope of the MCHF inaccuracy
oscillations at high magnetic field decreases linearly to 0 as a
function of $1/B$ and that in the strictly infinite magnetic field
limit the exact energy is obtained for broken-symmetry HF solution
with the classical point-charge distribution.
\newline

{\bf Acknowledgments} This work was supported in part by the
Polish Government Scientific Research Committee (KBN), the Flemish
Science Foundation (FWO-Vl), IUAP, and GOA. One of us (EA) is
supported by a EU Marie Curie fellowship. The first author (BS) is
supported by the Foundation for Polish Science (FNP).
\newline

\appendix{\bf Appendix}\newline

In order to derive wave function [Eq.~(\ref{wv})] let us first
consider the single electron in a homogeneous magnetic field. In
the Landau gauge, i.e., for $\mathbf{A}(\mathbf{r})=(-By,0,0)$,
the Hamiltonian has the form
\begin{equation}
H_0 = -\frac{\hbar^2}{2m^*} \left(\frac{\partial^2}{\partial x^2}
+\frac{\partial^2}{\partial y^2}\right) +i\hbar\omega_c
y\frac{\partial}{\partial x} +\frac{m^*\omega_c^2}{2} y^2 \;,
\label{H0}
\end{equation}
where $\omega_c=eB/m^*$. The lowest Landau energy level
$E_0=\hbar\omega_c/2$ is infinitely-fold degenerate. The
corresponding degenerate eigenstates have the form
\begin{equation}
\chi_q(x,y)=C_1 \exp[iqx-(\beta/2)(y-q/\beta)^2] \;, \label{chi}
\end{equation}
where $\beta=m^*\omega_c/\hbar=eB/\hbar$, $C_1$ is the
normalization constant, and $q \in (-\infty,+\infty)$. Due to this
degeneracy, an arbitrary linear combination of wave functions
(\ref{chi}) is an eigenfunction of Hamiltonian (\ref{H0}) to
eigenvalue $E_0$. The most general form of this linear combination
can be written as
\begin{equation}
\psi_0(x,y)=\int_{-\infty}^{+\infty} f(q)\chi_q(x,y) dq \; .
\label{psi1}
\end{equation}
Taking
\begin{equation}
f(q)=\exp\left[q(y_0-ix_0)-q^2/2\beta\right] \label{fq}
\end{equation}
and performing the integration in Eq.~(\ref{psi1}), we obtain
\begin{eqnarray}
\psi_0(x,y)&=C_0 \exp\{-(\beta /4)
\left[(x-x_0)^2+(y-y_0)^2\right] +\nonumber \\ &
(i\beta/2)(x-x_0)(y+y_0)\} \; , \label{psi2}
\end{eqnarray}
where $C_0$ is the normalization constant. Wave function
(\ref{psi2}) corresponds to the electron probability density,
which is localized around center $\mathbf{r}_0=(x_0,y_0)$. For
arbitrary $\mathbf{r}_0$ wave function (\ref{psi2}) is the
eigenfunction of the Hamiltonian (\ref{H0}) associated with
eigenvalue $E_0$. Due to the arbitrary choice of $\mathbf{r}_0$,
wave functions (\ref{psi2}), localized at different centers,
correspond to the same lowest Landau level.

If, in addition to the magnetic field, the electron is subject to
the external parabolic potential centered at site
$\mathbf{R}=(X,Y)$, i.e.,
\begin{equation}
V_{con\!f}(x,y)=\frac{m^*\omega_0^2}{2} [(x-X)^2+(y-Y)^2] \;,
\end{equation}
we deal with the Fock-Darwin eigenproblem. Then, the ground-state
wave function is centered around $\mathbf{R}$ and is written down
in the normalized form as
\begin{eqnarray}
\psi_{(X,Y)}&(x,y) =\exp\{-(\alpha /4) \left[(x-X)^2
+(y-Y)^2\right]+\nonumber \\ &
(i\beta/2)(x-X)(y+Y)\}/(\alpha/2\pi)^{1/2}, \label{psi3}
\end{eqnarray}
where $\alpha=(2m^*/\hbar)\sqrt{\omega_0^2+\omega_c^2/4}$.
Therefore, we obtain the wave function of form (\ref{wv}).

\end{document}